\documentstyle[fleqn,11pt]{book}
\begin{document}
\parindent 1.4cm
\large
\begin{center}
{{\bf THE QUANTUM WAVE PACKET AND THE FEYNMAN-DE BROGLIE-BOHM
PROPAGATOR OF THE LINEARIZED KOSTIN EQUATION ALONG A CLASSICAL
TRAJETORY}}
\end{center}
\begin{center}
{J. M. F. Bassalo$^{1}$,\ P. T. S. Alencar$^{2}$,\  D. G. da
Silva$^{3}$,\ A. B. Nassar$^{4}$\ and\ M. Cattani$^{5}$}
\end{center}
\begin{center}
{$^{1}$\ Funda\c{c}\~ao Minerva,\ Avenida Governador Jos\'e Malcher\ 629
-\ CEP\ 66035-100,\ Bel\'em,\ Par\'a,\ Brasil}
\end{center}
\begin{center}
{E-mail:\ jmfbassalo@gmail.com}
\end{center}
\begin{center}
{$^{2}$\ Universidade Federal do Par\'a\ -\ CEP\ 66075-900,\ Guam\'a,
Bel\'em,\ Par\'a,\ Brasil}
\end{center}
\begin{center}
{E-mail:\ tarso@ufpa.br}
\end{center}
\begin{center}
{$^{3}$\ Escola Munguba do Jari, Vit\'oria do Jari\ -\ CEP\
68924-000,\ Amap\'a,\ Brasil}
\end{center}
\begin{center}
{E-mail:\ danielgemaque@yahoo.com.br}
\end{center}
\begin{center}
{$^{4}$\ Extension Program-Department of Sciences, University of
California,\ Los Angeles, California 90024,\ USA}
\end{center}
\begin{center}
{E-mail:\ nassar@ucla.edu}
\end{center}
\begin{center}
{$^{5}$\ Instituto de F\'{\i}sica da Universidade de S\~ao Paulo. C. P.
66318, CEP\ 05315-970,\ S\~ao Paulo,\ SP, Brasil}
\end{center}
\begin{center}
{E-mail:\ mcattani@if.usp.br}
\end{center}
\par
{\underline {Abstract}}:\ In this paper we study the quantum wave
packet and the Feynman-de Broglie-Bohm propagator of the
linearized Kostin equation along a classical trajetory.
\vspace{0.2cm}
\par
PACS\ 03.65\ -\ Quantum Mechanics
\par
\vspace{0.2cm}
\par
1.\ {\bf Introduction}
\vspace{0.2cm}
\par
In the present work we investigate the quantum wave packet and the
Feynman-de Broglie-Bohm propagator of the linearized Kostin
equation along a classical trajetory by using the
\vspace{0.2cm}
\par
2.\ {\bf The Kostin Equation}
\par
Em 1972,\ [1] M. D. Kostin proposed a non-linear Schr\"{o}dinger to
represent time dependent physical systems, given by:
\begin{center}
{$i\ {\hbar}\ {\frac {{\partial}}{{\partial}t}}\ {\psi}(x,\ t)\ =\ -\
{\frac {{\hbar}^{2}}{2\ m}}\ {\frac {{\partial}^{2}{\psi}(x,\
t)}{{\partial}x^{2}}}\ +$}
\end{center}
\begin{center}
{$+\ [V(x,\ t)\ +\ {\frac {{\hbar}\ {\nu}}{2\ i}}\ {\ell}n\ {\frac
{{\psi}(x,\ t)}{{\psi}^{*}(x,\ t)}}]\ {\psi}(x,\ t)$\ ,\ \ \ \ \
(2.1)}
\end{center}
where ${\psi}(x,\ t)$ and $V(x,\ t)$ are, respectively, the
wavefunction and the time dependent potential of the physical system in
study, and ${\nu}$ is a constant.
\par
Writting the wavefunction ${\psi}(x,\ t)$ in the polar form, defined by
the Madelung-Bohm [2,\ 3]:
\begin{center}
{${\psi}(x,\ t)\ =\ {\phi}(x,\ t)\ exp\ [i\ S(s,\ t)]$,\ \ \ \ \ (2.2)}
\end{center}
where $S(x,\ t)$ is the classical action and ${\phi}(x,\ t)$ will be
defined in what follows, and using eq. (2.2) in eq. (2.1), we get:\ [4]
\begin{center}
{$i\ {\hbar}\ (i\ {\frac {{\partial}S}{{\partial}t}}\ +\ {\frac
{1}{{\phi}}}\ {\frac {{\partial}{\phi}}{{\partial}t}})\ {\psi}\ =$}
\end{center}
\begin{center}
{$=\ -\ {\frac {{\hbar}^{2}}{2\ m}}\ [i\ {\frac
{{\partial}^{2}S}{{\partial}x^{2}}}\ +\ {\frac {1}{{\phi}}}\ {\frac
{{\partial}^{2}{\phi}}{{\partial}x^{2}}}\ -\ ({\frac
{{\partial}S}{{\partial}x}})^{2}\ +\ 2\ {\frac {i}{{\phi}}}\ {\frac
{{\partial}S}{{\partial}x}}\ {\frac {{\partial}{\phi}}{{\partial}x}}]\
{\psi}\ +$}
\end{center}
\begin{center}
{$+\ [V(x,\ t)\ +\ {\frac {{\hbar}\ {\nu}}{2\ i}}\ {\ell}n\ {\frac
{{\phi}\ e^{i\ S}}{{\phi}\ e^{-\ i\ S}}}]\ {\psi}\ \ \ {\to}$}
\end{center}
\begin{center}
{$i\ {\hbar}\ (i\ {\frac {{\partial}S}{{\partial}t}}\ +\ {\frac
{1}{{\phi}}}\ {\frac {{\partial}{\phi}}{{\partial}t}})\ {\psi}\ =$}
\end{center}
\begin{center}
{$=\ -\ {\frac {{\hbar}^{2}}{2\ m}}\ [i\ {\frac
{{\partial}^{2}S}{{\partial}x^{2}}}\ +\ {\frac {1}{{\phi}}}\ {\frac
{{\partial}^{2}{\phi}}{{\partial}x^{2}}}\ -\ ({\frac
{{\partial}S}{{\partial}x}})^{2}\ +\ 2\ {\frac {i}{{\phi}}}\ {\frac
{{\partial}S}{{\partial}x}}\ {\frac {{\partial}{\phi}}{{\partial}x}}]\
{\psi}\ +$}
\end{center}
\begin{center}
{$+\ [V(x,\ t)\ +\ {\hbar}\ {\nu}\ S]\ {\psi}$\ .\ \ \ \ \ (2.3)}
\end{center}
\par
Taking the real and imaginary parts of eq. (2.3), we obtain:
\par
a)\ {\underline {imaginary part}}
\begin{center}
{${\frac {{\hbar}}{{\phi}}}\ {\frac {{\partial}{\phi}}{{\partial}t}}\
=\ -\ {\frac {{\hbar}^{2}}{2\ m}}\ ({\frac
{{\partial}^{2}S}{{\partial}x^{2}}}\ +\ {\frac {2}{{\phi}}}\ {\frac
{{\partial}S}{{\partial}x}}\ {\frac {{\partial}{\phi}}{{\partial}x}})\
\ \ {\to}$}
\end{center}
\begin{center}
{${\frac {{\partial}{\phi}}{{\partial}t}}\ =\ -\ {\frac {{\hbar}}{2\
m}}\ ({\phi}\ {\frac {{\partial}^{2}S}{{\partial}x^{2}}}\ +\ 2\ {\frac
{{\partial}S}{{\partial}x}}\ {\frac {{\partial}{\phi}}{{\partial}x}})$\
,\ \ \ \ \ (2.4)}
\end{center}
\par
b)\ {\underline {real part}}
\begin{center}
{$-\ {\hbar}\ {\frac {{\partial}S}{{\partial}t}}\ =\ -\ {\frac
{{\hbar}^{2}}{2\ m}}\ [{\frac {1}{{\phi}}}\ {\frac
{{\partial}^{2}{\phi}}{{\partial}x^{2}}}\ -\ ({\frac
{{\partial}S}{{\partial}x}})^{2}]\ +$}
\end{center}
\begin{center}
{$+\ [V(x,\ t)\ +\ {\hbar}\ {\nu}\ S]\ \ \ \ \ {\to}$}
\end{center}
\begin{center}
{$-\ {\frac {{\hbar}}{m}}\ {\frac {{\partial}S}{{\partial}t}}\ =\ -\ {\frac
{{\hbar}^{2}}{2\ m^{2}}}\ {\frac {1}{{\phi}}}\ [{\frac
{{\partial}^{2}{\phi}}{{\partial}x^{2}}}\ -\ {\phi}\ ({\frac
{{\partial}S}{{\partial}x}})^{2}]\ +$}
\end{center}
\begin{center}
{$+\ {\frac {1}{m}}\ [V(x,\ t)\ +\ {\hbar}\ {\nu}\ S]$\ .\ \ \ \ \
(2.5)}
\end{center}
\par
\vspace{0.2cm}
2.1 {\bf Dynamics of the Kostin Equation}
\par
Now, let us see the correlation between eqs. (2.4,5) and the
traditional equations of the Real Fluid Dynamics:\ a)\ continuity
equation and b)\ Navier-Stokes's equation. To do is let us perform the
following correspondences:
\begin{center}
{${\sqrt {{\rho}(x,\ t)}}\ =\ {\phi}(x,\ t)$\ ,\ \ \ \ \ (2.6)\ \ \
(quantum mass density)}
\end{center}
\begin{center}
{$v_{qu}(x,\ t)\ =\ {\frac {{\hbar}}{m}}\ {\frac {{\partial}S(x,\
t)}{{\partial}x}}$\ ,\ \ \ \ \ (2.7)\ \ \ \ \ (quantum velocity)}
\end{center}
\begin{center}
{$V_{qu}(x,\ t)\ =\ -\ {\frac {{\hbar}^{2}}{2\ m}}\ {\frac {1}{{\sqrt
{{\rho}}}}}\ {\frac {{\partial}^{2}{\sqrt
{{\rho}}}}{{\partial}x^{2}}}\ =\ -\ {\frac {{\hbar}^{2}}{2\ m\
{\phi}}}\ {\frac {{\partial}^{2}{\phi}}{{\partial}x^{2}}}$\
.\ \ \ \ \ (2.8a,b)\ \ \ \ \ (Bohm quantum potential)}
\end{center}
\par
Putting eq. (2.6,7) into (2.4) we get:
\begin{center}
{${\frac {{\partial}{\sqrt {{\rho}}}}{{\partial}t}}\ =\ -\ {\frac
{{\hbar}}{2\ m}}\ {\big {(}}\ 2\ {\frac {{\partial}S}{{\partial}x}}\
{\frac {{\partial}{\sqrt {{\rho}}}}{{\partial}x}}\ +\ {\sqrt {{\rho}}}\
{\frac {{\partial}^{2}S}{{\partial}x^{2}}}{\big {)}}\ \ \ \ \ {\to}$}
\end{center}
\begin{center}
{${\frac {1}{2\ {\sqrt {{\rho}}}}}\ {\frac
{{\partial}{\rho}}{{\partial}t}}\ =\ -\ {\frac {{\hbar}}{2\ m}}\ {\big
{(}}\ 2\ {\frac {{\partial}S}{{\partial}x}}\ {\frac {1}{2\ {\sqrt
{{\rho}}}}}\ {\frac {{\partial}{\rho}}{{\partial}x}}\ +\ {\sqrt
{{\rho}}}\ {\frac {{\partial}^{2}S}{{\partial}x^{2}}}\ {\big {)}}\ \ \
\ \ {\to}$}
\end{center}
\begin{center}
{${\frac {1}{{\rho}}}\ {\frac
{{\partial}{\rho}}{{\partial}t}}\ =\ -\ {\frac {{\hbar}}{m}}\ {\big
{(}}\ {\frac {{\partial}S}{{\partial}x}}\ {\frac {1}{{\rho}}}\
{\frac {{\partial}{\rho}}{{\partial}x}}\ +\ {\frac
{{\partial}^{2}S}{{\partial}x^{2}}}\ {\big {)}}\ \ \ \ \ {\to}$}
\end{center}
\begin{center}
{${\frac {1}{{\rho}}}\ {\frac
{{\partial}{\rho}}{{\partial}t}}\ =\ -\ {\frac
{{\partial}}{{\partial}x}}\ {\big {(}}\ {\frac {{\hbar}}{m}}\ {\frac
{{\partial}S}{{\partial}x}}\ {\big {)}}\ -\ {\frac {1}{{\rho}}}\ {\big
{(}}\ {\frac {{\hbar}}{m}}\ {\frac {{\partial}S}{{\partial}x}}\ {\big
{)}}\ {\frac {{\partial}{\rho}}{{\partial}x}}\ \ \ \ \ {\to}$}
\end{center}
\begin{center}
{${\frac {1}{{\rho}}}\ {\frac
{{\partial}{\rho}}{{\partial}t}}\ =\ -\ {\frac
{{\partial}v_{qu}}{{\partial}x}}\ -\ {\frac {v_{qu}}{{\rho}}}\ {\frac
{{\partial}{\rho}}{{\partial}x}}\ \ \ \ \ {\to}$}
\end{center}
\begin{center}
{${\frac {{\partial}{\rho}}{{\partial}t}}\ +\ {\rho}\ {\frac
{{\partial}v_{qu}}{{\partial}x}}\ +\ v_{qu}\ {\frac
{{\partial}{\rho}}{{\partial}x}}\ =\ 0\ \ \ \ \ {\to}$}
\end{center}
\begin{center}
{${\frac {{\partial}{\rho}}{{\partial}t}}\ +\ {\frac
{{\partial}({\rho}\ v_{qu})}{{\partial}x}}\ =\ 0$\ ,\ \ \ \ \ (2.9)}
\end{center}
which represents the continuity equation of the mass conservation law
of the Fluid Dynamics. We must note that this expression also
indicates descoerence of the considered physical system represented by
(2.1).
\par
Now, taking the eq. (2.5) and using the eqs. (2.7,8b), will be:
\begin{center}
{$-\ {\hbar}\ {\frac {{\partial}S}{{\partial}t}}\ =\ -\ ({\frac
{{\hbar}^{2}}{2\ m\ {\phi}}})\ {\frac
{{\partial}^{2}{\phi}}{{\partial}x^{2}}}\ +\ {\frac {1}{2}}\ m\ ({\frac
{{\hbar}}{m}}\ {\frac {{\partial}S}{{\partial}x}})^{2}\ +$}
\end{center}
\begin{center}
{$+\ [V(x,\ t)\ +\ {\hbar}\ {\nu}\ S]\ \ \ {\to}$}
\end{center}
\begin{center}
{${\hbar}\ ({\frac {{\partial}S}{{\partial}t}}\ +\ {\nu}\ S)\ +\
({\frac {1}{2}}\ m\ v_{qu}^{2}\ +\ V\ +\ V_{qu})\ =\ 0$\ .\ \ \ \ \
(2.10)}
\end{center}
\par
Differentiating the eq. (2.5) with respect $x$ and using the eqs. (2.7,8b)
we have:
\begin{center}
{$-\ {\frac {{\hbar}}{m}}\ {\frac {{\partial}^{2}S}{{\partial}x\
{\partial}t}}\ =\ -\ {\frac {{\hbar}^{2}}{2\ m^{2}}}\ {\frac
{{\partial}}{{\partial}x}}\ [{\frac {1}{{\phi}}}\ {\frac
{{\partial}^{2}{\phi}}{{\partial}x^{2}}}\ -\ ({\frac
{{\partial}S}{{\partial}x}})^{2}]\ +\ {\frac {1}{m}}\ {\frac
{{\partial}V}{{\partial}x}}\ +\ {\frac {{\hbar}}{m}}\ {\nu}\ {\frac
{{\partial}S}{{\partial}x}}\ \ \ {\to}$}
\end{center}
\begin{center}
{$-\ {\frac {{\partial}}{{\partial}t}}\ ({\frac {{\hbar}}{m}}\ {\frac
{{\partial}S}{{\partial}x}})\ =\ {\frac {1}{m}}\ {\frac
{{\partial}}{{\partial}x}}\ (-\ {\frac {{\hbar}^{2}}{2\ m}}\ {\frac
{1}{{\phi}}}\ {\frac {{\partial}^{2}{\phi}}{{\partial}x^{2}}})$\ +}
\end{center}
\begin{center}
{+\ ${\frac {1}{2}}\ {\frac {{\partial}}{{\partial}x}}\ ({\frac
{{\hbar}}{m}}\ {\frac {{\partial}S}{{\partial}x}})^{2}\ +\
{\frac {1}{m}}\ {\frac {{\partial}V}{{\partial}x}}\ +\ {\nu}\ {\frac
{{\hbar}}{m}}\ {\frac {{\partial}S}{{\partial}x}}\ \ \ {\to}$}
\end{center}
\begin{center}
{${\frac {{\partial}v_{qu}}{{\partial}t}}\ +\ v_{qu}\ {\frac
{{\partial}v_{qu}}{{\partial}x}}\ +$}
\end{center}
\begin{center}
{$+\ {\nu}\ v_{qu}\ = -\ {\frac {1}{m}}\ {\frac
{{\partial}}{{\partial}x}}\ (V\ +\ V_{qu})$\ ,\ \ \ \ \ (2.11)}
\end{center}
which is an equation similar to the Navier-Stokes's equation which
governs the motion of an real fluid.
\par
Considering the "substantive differentiation" (local plus convective)
or "hidrodynamic differention":\ $d/dt\ =\ {\partial}/{\partial}t\ +
v_{qu}\ {\partial}/{\partial}x$ and that $v_{qu}\ =\ dx_{qu}/dt$, the
eq. (2.11) could be written as:[5]
\begin{center}
{$m\ d^{2}x/dt^{2}\ =\ -\ {\nu}\ v_{qu}\ -\ {\frac {1}{m}}\ {\frac
{{\partial}}{{\partial}x}}\ (V\ +\ V_{qu})$\ ,\ \ \ \ \ (2.12)}
\end{center}
that has a form of the $second\ Newton\ law$.
\vspace{0.2cm}
\par
3.\ {\bf The Quantum Wave Packet of the Linearized Kostin
Equation along a Classical Trajetory} \vspace{0.2cm}
\par
In order to find the quantum wave packet of the non-linear Kostin
equation, let us considerer the following $ansatz$:[6]
\par
\begin{center}
{${\rho}\ (x,\ t) =\ [2{\pi}\ a^{2}(t)]^{-\ 1/2}\ exp\ {\Big {(}}\ -\
{\frac {[x\ -\ q(t)]^{2}}{2\ a^{2}(t)}}\ {\Big {)}}$,\ \ \ \ \ (3.1)}
\end{center}
where $a(t)$ and $q(t)$ are auxiliary functions of time, to be
determined in what follows;\ they represent the {\it width} and {\it center
of mass of wave packet}, respectively.
\par
Substituting (3.1) into (2.3) and integrated the result, we
obtain:[4]
\begin{center}
{$v_{qu}\ (x,\ t)\ =\ {\frac {{\dot {a}}(t)}{a(t)}}\ [x\ -\ q(t)]\ +\
{\dot {q}}(t)$,\ \ \ \ \ (3.2)}
\end{center}
where the integration constant must be equal to zero since ${\rho}$ and
${\rho}\ {\partial {S}}/{\partial {x}}$ vanish for ${\mid}\ x\ {\mid}\
\ \ {\to}\ \ \ {\infty}$.\ In fact, any well-behaved function of (x\ -\
X) multiplied by ${\rho}$ clearly vanishes as ${\mid}\ x\ {\mid}\
\ \ {\to}\ \ \ {\infty}$.
\par
To obtain the quantum wave packet of the linear Kostin equation
along a classical trajetory given by (2.1), let us expand the
functions $S(X,\ T)$, $V(x,\ t)$ and $V_{qu}(x,\ t)$ around of
$q(t)$ up to second Taylor order.\ In this way we have:
\begin{center}
{$S(x,\ t)\ =\ S[q(t),\ t]\ +\ S'[q(t),\ t]\ [x\ -\ q(t)]\ +\ {\frac
{S''[q(t),\ t]}{2}}\ [x\ -\ q(t)]^{2}$\ ,\ \ \ \ \ (3.3)}
\end{center}
\begin{center}
{$V(x,\ t)\ =\ V[q(t),\ t]\ +\ V'[q(t),\ t]\ [x\ -\
q(t)]\ +\ {\frac {V''[q(t),\ t]}{2}}\ [x\ -\ q(t)]^{2}$\ .\ \ \ \ \ (3.4)}
\end{center}
\begin{center}
{$V_{qu}(x,\ t)\ =\ V_{qu}[q(t),\ t]\ +\ V_{qu}'[q(t),\ t]\ [x\ -\
q(t)]\ +\ {\frac {V_{qu}''[q(t),\ t]}{2}}\ [x\ -\ q(t)]^{2}$\ .\ \ \ \
\ (3.5)}
\end{center}
\par
Differentiating (3.3) in the variable $x$, multiplying
the result by ${\frac {{\hbar}}{m}}$, using the eqs. (2.7) and
(3.2), taking into account the polynomial identity property and also
considering the second Taylor order, we obtain:
\begin{center}
{${\frac {{\hbar}}{m}}\ {\frac {{\partial}S(x,\ t)}{{\partial}x}}\ =\
{\frac {{\hbar}}{m}}\ {\Big {(}}\ S'[q(t),\ t]\ +\ S''[q(t),\ t]\ [x\
-\ q(t)]\ {\Big {)}}\ =$}
\end{center}
\begin{center}
{$=\ v_{qu}(x,\ t)\ =\ {\big {[}}\ {\frac {\dot {a}(t)}{a(t)}}\ {\big
{]}}\ [x_{qu}\ -\ q(t)]\ +\ {\dot {q}}(t)\ \ \ \ {\to}$}
\end{center}
\begin{center}
{$S'[q(t),\ t]\ =\ {\frac {m\ {\dot {q}}(t)}{{\hbar}}}\ ,\ \ \
S''[q(t),\ t]\ =\ {\frac {m}{{\hbar}}}\ {\frac {{\dot
{a}}(t)}{a(t)}}$\ ,\ \ \ \ \ (3.6a,b)}
\end{center}
\par
Substituting (3.6a,b) into (3.3), results:
\begin{center}
{$S(x,\ t)\ =\ S_{o}(t)\ +\ {\frac {m\ {\dot {q}}(t)}{{\hbar}}}\ [x\ -\
q(t)]\ +\ {\frac {m}{2\ {\hbar}}}\ {\frac {{\dot {a}}(t)}{a(t)}}\ [x\
-\ q(t)]^{2}$\ ,\ \ \ \ \ (3.7)}
\end{center}
where:
\begin{center}
{$S_{o}(t)\ {\equiv}\ S[q(t),\ t]$\ ,\ \ \ \ \ (3.8)}
\end{center}
are the classical actions.
\par
Differentiating the (3.7) with respect to $t$, we obtain
(remembering that ${\frac {{\partial}x}{{\partial}t}}$\ =\ 0):
\begin{center}
{${\frac {{\partial}S}{{\partial}t}}\ =\ {\dot {S}}_{o}(t)\ +\ {\frac
{{\partial}}{{\partial}t}}\ {\Big {(}}\ {\frac {m\ {\dot
{q}}(t)}{{\hbar}}}\ [x\ -\ q(t)]\ {\Big {)}}\ +\ {\frac
{{\partial}}{{\partial}t}}\ {\Big {(}}\ {\frac {m}{2\ {\hbar}}}\ {\Big
{[}}\ {\frac {{\dot {a}}(t)}{a(t)}}\ {\Big {]}}\ [x\ -\ q(t)]^{2}\
{\Big {)}}\ \ \ {\to}$}
\end{center}
\begin{center}
{${\frac {{\partial}S}{{\partial}t}}\ =\ {\dot {S}}_{o}(t)\ +\ {\frac
{m\ {\ddot {q}}(t)}{{\hbar}}}\ [x\ -\ q(t)]\ -\ {\frac {m\ {\dot
{q}}(t)^{2}}{{\hbar}}}\ +$}
\end{center}
\begin{center}
{+\ ${\frac {m}{2\ {\hbar}}}\ [{\frac {{\ddot
{a}}(t)}{a(t)}}\ -\ {\frac {{\dot
{a}}^{2}(t)}{a^{2}(t)}}]\ [x\ -\ q(t)]^{2}\ -\ {\frac {m\
{\dot {q}}(t)}{{\hbar}}}\ {\frac {{\dot
{a}}(t)}{a(t)}}\ [x\ -\ q(t)]$\ .\ \ \ \ \ (3.9)}
\end{center}
\par
Considering the eqs. (2.6) and (3.1), let us write $V_{qu}$ given by
(2.8a,b) in terms of potencies of $[x\ -\ q(t)]$. Initially using (2.5)
and (3.1), we calculate the following derivations:
\begin{center}
{${\frac {{\partial}{\phi}}{{\partial}x}}\ =\ {\frac
{{\partial}}{{\partial}x}}\ {\Big {(}}\ [2\ {\pi}\ a^{2}(t)]^{-\
1/4}\ e^{-\ {\frac {[x\ -\ q(t)]^{2}}{4\ a^{2}(t)}}}\ {\Big
{)}}\ =\ [2\ {\pi}\ a^{2}(t)]^{-\ 1/4}\ e^{-\ {\frac {[x\ -\
q(t)]^{2}}{4\ a^{2}(t)}}} {\frac {{\partial}}{{\partial}x}}\
{\Big {(}}\ -\ {\frac {[x\ -\ q(t)]^{2}}{4\ a^{2}(t)}}\ {\Big
{)}}\ \ \ {\to}$}
\end{center}
\begin{center}
{${\frac {{\partial}{\phi}}{{\partial}x}}\ =\ -\ [2\ {\pi}\
a^{2}(t)]^{-\ 1/4}\ e^{-\ {\frac {[x\ -\ q(t)]^{2}}{4\
^{2}(t)}}}\ {\frac {[x\ -\ q(t)]}{2\ a^{2}(t)}}$\ ,}
\end{center}
\begin{center}
{${\frac {{\partial}^{2}{\phi}}{{\partial}x^{2}}}\ =\ {\frac
{{\partial}}{{\partial}x}}\ {\Big {(}}\ -\ [2\ {\pi}\
a^{2}(t)]^{-\ 1/4}\ e^{-\ {\frac {[x\ -\ q(t)]^{2}}{4\
a^{2}(t)}}}\ {\frac {[x\ -\ q(t)]}{2\ a^{2}(t)}}\ {\Big
{)}}$\ =}
\end{center}
\begin{center}
{$\ =\ -\ [2\ {\pi}\ a^{2}(t)]^{-\ 1/4}\ e^{-\ {\frac {[x\ -\
q(t)]^{2}}{4\ a^{2}(t)}}}\ {\frac {{\partial}}{{\partial}x}}\
{\Big {(}}\ {\frac {[x\ -\ q(t)]}{2\ a^{2}(t)}}\ {\Big {)}}\ -$}
\end{center}
\begin{center}
{$-\ [2\ {\pi}\ a^{2}(t)]^{-\ 1/4}\ e^{-\ {\frac {[x\ -\
q(t)]^{2}}{4\ a^{2}(t)}}}\ {\frac {{\partial}}{{\partial}x}}\
{\Big {(}}\ -\ {\frac {[x\ -\ q(t)]^{2}}{4\ a^{2}(t)}}\ {\Big
{)}}\ \ \ {\to}$}
\end{center}
\begin{center}
{${\frac {{\partial}^{2}{\phi}}{{\partial}x^{2}}}\ =\ -\ [2\ {\pi}\
a^{2}(t)]^{-\ 1/4}\ e^{-\ {\frac {[x\ -\ q(t)]^{2}}{4\
a^{2}(t)}}}\ {\frac {1}{2\ a^{2}(t)}}\ +\ [2\ {\pi}\
a^{2}(t)]^{-\ 1/4}\ e^{-\ {\frac {[x\ -\ q(t)]^{2}}{4\
a^{2}(t)}}}\ {\frac {[x\ -\ q(t)]^{2}}{4\ a^{4}(t)}}$\ =}
\end{center}
\begin{center}
{$=\ -\ {\phi}\ {\frac {1}{2\ a^{2}(t)}}\ +\ {\phi}\ {\frac {[x\
-\ q(t)]^{2}}{4\ a^{4}(t)}}\ \ \ {\to}\ \ \ {\frac {1}{{\phi}}}\
{\frac {{\partial}^{2}{\phi}}{{\partial}x^{2}}}\ =\ {\frac {[x\ -\
q(t)]^{2}}{4\ a^{4}(t)}}\ -\ {\frac {1}{2\ a^{2}(t)}}$\
.\ \ \ \ \ (3.10)}
\end{center}
\par
Substituting (3.10) into (2.8a) and taking into account (3.5), results:
\begin{center}
{$V_{qu}(x,\ t)\ =\ V_{qu}[q(t),\ t]\ +\ V_{qu}'[q(t),\ t]\ [x\ -\
q(t)]\ +\ {\frac {V_{qu}''[q(t),\ t]}{2}}\ [x\ -\ q(t)]^{2}\ \ \ {\to}$}
\end{center}
\begin{center}
{$V_{qu}(x,\ t)\ =\ {\frac {{\hbar}^{2}}{4\ m\ a^{2}(t)}}\ -\ {\frac
{{\hbar}^{2}}{8\ m\ a^{4}(t)}}\ [x\ -\ q(t)]^{2}$\ .\ \ \ \ \ (3.11)}
\end{center}
\par
Inserting the eqs. (2.7), (3.2,3,4), (3.7,8,9) and (3.11), into
(2.11), we obtain [remembering that $S_{o}(t)$, $a(t)$ and $q(t)$]:
\begin{center}
{${\hbar}\ [{\frac {{\partial}S}{{\partial}t}}\ +\ {\nu}\ S\ +\
({\frac {1}{2}}\ m\ v_{qu}^{2}\ +\ V\ +\ V_{qu})\ =\ 0$.\ \ \ \ \ \ (3.12)}
\end{center}
\begin{center}
{$\ {\hbar}\ {\Big {(}}\ {\dot {S}}_{o}\ +\ {\frac {m\ {\ddot
{q}}(t)}{{\hbar}}}\ [x\ -\ q(t)]\ -\ {\frac {m\ {\dot
{q}}^{2}(t)}{{\hbar}}}\ +\ {\frac {m}{2\ {\hbar}}}\ {\big {[}}\ {\frac
{{\ddot {a}}(t)}{a(t)}}\ -\ {\frac {{\dot
{a}}^{2}}{a^{2}(t)}}\ {\big {]}}\ [x\ -\ q(t)]^{2}\ -$}
\end{center}
\begin{center}
{$-\ {\frac {m\ {\dot {q}}(t)}{{\hbar}}}\ {\frac {{\dot
{a}}(t)}{a}(t)}\ [x\ -\ q(t)]\ {\Big {)}}\ +\ {\frac {m}{2}}\
{\Big {(}}\ {\frac{{\dot {a}}(t)}{a(t)}}\ [x\ -\ q(t)]\ +\ {\dot
{q}}(t)\ {\Big {)}}^{2}\ +$}
\end{center}
\begin{center}
{$+\ {\hbar}\ {\nu}\ {\bigg {(}}\ S_{0}(t)\ +\ {\frac {m\ {\dot
{q}}(t)}{{\hbar}}}\ [x\ -\ q(t)]\ +\ {\frac {m}{2\ {\hbar}}}\ {\frac
{{\dot {a}}(t)}{a(t)}}\ [x\ -\ q(t)]^{2}\ {\bigg {)}}\ +$}
\end{center}
\begin{center}
{$+\ V[q(t),\ t]\ +\ V'[q(t),\ t]\ [x\ -\ q(t)]\ +\ {\frac {m}{2}}\ V''[q(t),\
t]\ [x\ -\ q(t)]^{2}$\ +}
\end{center}
\begin{center}
{$+\ {\frac {{\hbar}^{2}}{4\ m\ a^{2}(t)}}\ -\ {\frac
{{\hbar}^{2}}{8\ m\ a^{4}(t)}}\ [x\ -\ q(t)]^{2}\ =\ 0$\ .\ \ \ \ \
(3.13)}
\end{center}
\par
Since $(x\ -\ q)^{o}\ =\ 1$, expanding (3.13) in potencies of $(x\ -\
q)$, we obtain:
\begin{center}
{${\Bigg {(}}\ {\hbar}\ {\dot {S}}_{o}(t)\ -\ m\ {\dot {q}}^{2}(t)\ +\
{\frac {1}{2}}\ m\ {\dot {q}}^{2}(t)\ +\ V[q(t),\ t]\ +\ {\hbar}\
{\nu}\ [S_{0}(t)]\ +\ {\frac {{\hbar}^{2}}{4\ m\ a^{2}(t)}}\ {\Bigg
{)}}\ [x\ -\ q(t)]^{o}\ +$}
\end{center}
\begin{center}
{$+\ {\Big {(}}\ m\ {\ddot {q}}(t)\ -\ m\ {\dot {q}}(t)\ {\frac
{{\dot {a}}(t)}{a(t)}}\ +\ m\ {\dot {q}}(t)\ {\frac
{{\dot {a}}(t)}{a(t)}}\ +\ {\nu}\ m\ {\dot {q}}(t)\ +\ V'[q(t),\ t]\
{\Big {)}}\ [x\ -\ q(t)]$\ +}
\end{center}
\begin{center}
{$+\ {\Big {(}}\ {\frac {m}{2}}\ [{\frac {{\ddot
{a}}(t)}{a(t)}}\ -\ {\frac {{\dot {a}}(t)^{2}}{a(t)^{2}}}]\ +\ {\frac
{m}{2}}\ {\frac {{\dot {a}}^{2}(t)}{a^{2}(t)}}\ +\ m\ {\nu}\ {\frac
{{\dot {a}}(t)}{a(t)}}\ +$}
\end{center}
\begin{center}
{$+\ {\frac {1}{2}}\ V"[q(t),\ t]\ -\ {\frac
{{\hbar}^{2}}{4\ m\ a^{4}(t)}}\ {\Big {)}}\ [x\ -\ q(t)]^{2}\ =\ 0$\ .\
\ \ \ \ (3.14)}
\end{center}
\par
As (3.14) is an identically null polynomium, all coefficients of the
potencies must be all equal to zero, that is:
\begin{center}
{${\dot {S}}_{o}(t)\ =\ {\frac {1}{{\hbar}}}\ {\Bigg {(}}\ {\frac
{1}{2}}\ m\ {\dot {q}}(t)^{2}\ -\ V[q(t),\ t]\ -\ {\frac
{{\hbar}^{2}}{4\ m\ a(t)^{2}}}\ -\ {\hbar}\ {\nu}\ S_{0}(t)\ {\Bigg
{)}}$\ ,\ \ \ \ \ (3.15)}
\end{center}
\begin{center}
{${\ddot {q}}(t)\ +\ {\nu}\ {\dot {q}}(t)\ +\ {\frac {1}{m}}\
V'[q(t),\ t]\ =\ 0$\ ,\ \ \ \ \ (3.16)}
\end{center}
\begin{center}
{${\ddot {a}}(t)\ +\ {\nu}\ {\dot {a}}(t)]\ +\ {\frac {1}{m}}\
V"[q(t),\ t\ =\ {\frac {{\hbar}^{2}}{4\ m^{2}\ a^{3}(t)}}$\ .\ \ \ \ \
(3.17)}
\end{center}
\par
Assuming that the following initial conditions are obeyed:
\begin{center}
{$q(0)\ =\ x_{o}\ ,\ \ \ {\dot {q}}(0)\ =\ v_{o}\ ,\ \ \ a(0)\
=\ a_{o}\ ,\ \ \ {\dot {a}}(0)\ =\ b_{o}$\ ,\ \ \ \ \ \ (3.18a-d)}
\end{center}
and that:
\begin{center}
{$S_{o}(0)\ =\ {\frac {m\ v_{o}\ x_{o}}{{\hbar}}}$\ ,\ \ \ \ \ (3.19)}
\end{center}
the integration of (3.15) gives:
\begin{center}
{$S_{o}(t)\ =\ {\frac {1}{{\hbar}}}\ {\int}_{o}^{t}\ dt'\ {\Bigg {(}}\
{\frac {1}{2}}\ m\ {\dot {q}}^{2}(t')\ +\ {\nu}\ {\Big {[}}\
S_{0}(t')\ {\Big {]}}\ -\ $}
\end{center}
\begin{center}
{$-\ V[q(t'),\ t']\ -\ {\frac {{\hbar}^{2}}{4\ m\ a^{2}(t')}}\ {\Bigg
{)}}\ +\ {\frac {m\ v_{o}\ x_{o}}{{\hbar}}}$.\ \ \ \ \ (3.20)}
\end{center}
\par
Taking the eq. (3.20) in the eq. (3.7) results:
\begin{center}
{$S(x,\ t)\ =\ {\frac {1}{{\hbar}}}\ {\int}_{o}^{t}\ dt'\ {\Bigg {(}}\
{\frac {1}{2}}\ m\ {\dot {q}}^{2}(t')\ +\ {\hbar}\ {\nu}\
[S_{0}(t')]\ -\ V[q(t'),\ t']\ -\ {\frac
{{\hbar}^{2}}{4\ m\ a^{2}(t')}}\ {\Bigg {)}}\ +$}
\end{center}
\begin{center}
{$+\ {\frac {m\ v_{o}\ x_{o}}{{\hbar}}}\ +\ {\frac {m\ {\dot
{q}}(t)}{{\hbar}}}\ [x\ -\ q(t)]\ +\ {\frac {m}{2\ {\hbar}}}\ {\Big
{[}}\ {\frac {{\dot {a}}(t)}{a(t)}}\ [x\ -\ q(t)]^{2}\ {\Big {]}}$\ .\
\ \ \ \ (3.21)}
\end{center}
\par
The above result permit us, finally, to obtain the wave packet
for the linearized Kostin equation along a classical trajetory.
Indeed, considering (2.2), (2.6), (3.1) and (3.21), we get:\ [6]
\begin{center}
{${\psi}(x,\ t)\ =\ [2\ {\pi}\ a^{2}(t)]^{-\ 1/4}\ exp\ {\Bigg
{[}}\ {\frac {i\ m}{2\ {\hbar}}}\ {\Big {(}}\ {\frac {{\dot
{a}}(t)}{a(t)}}\ - {\frac {1}{4\ a^{2}(t)}}\ {\Big {)}}\ [x\ -\
q(t)]^{2}\ {\Bigg {]}}\ {\times}$}
\end{center}
\begin{center}
{${\times}\ exp\ {\Bigg {[}}\ {\frac {i\ m\ {\dot {q}}(t)}{{\hbar}}}\ [x\
-\ q(t)]\ +\ {\frac {i\ m\ v_{o}\ x_{o}}{{\hbar}}}\ {\Big {]}}\ {\times}$}
\end{center}
\begin{center}
{${\times}\ exp\ {\Bigg {[}}\ {\frac {i}{{\hbar}}}\ {\int}_{o}^{t}\ dt'\
{\Bigg {(}}\ {\frac {1}{2}}\ m\ {\dot {q}}^{2}(t')\ +\ {\hbar}\
{\nu}\ [S_{0}(t')\ -\ V[q(t'),\ t']\ -\ {\frac {{\hbar}^{2}}{4\ m\
a^{2}(t')}}\ {\Bigg {)}}\ {\Bigg {]}}$\ .\ \ \ \ \ (3.22)}
\end{center}
\par
Note that putting ${\nu}\ =\ 0$ into (3.22) we obtain the
quantum wave packet of the Schr\"{o}dinger equation with the potential
V(x,\ t).\ [7]
\par
\par
4.\ {\bf The Feynman-de Broglie-Bohm Propagator of the Linearized
Kostin Equation along a Classical Trajetory}
\par
\vspace{0.2cm}
\par
4.1.\ {\bf Introduction}
\vspace{0.2cm}
\par
In 1948,\ [8] Feynman formulated the following principle of minimum
action for the Quantum Mechanics:
\begin{center}
{{\it The transition amplitude between the states ${\mid}\ a\ >$ and
${\mid}\ b\ >$ of a quantum-mechanical system is given by the sum of
the elementary contributions, one for each trajectory passing by
${\mid}\ a\ >$ at the time t$_{a}$ and by ${\mid}\ b\ >$ at the time
t$_{b}$. Each one of these contributions have the same modulus, but its
phase is the classical action S$_{c{\ell}}$ for each trajectory.}}
\end{center}
\par
This principle is represented by the following expression known as the
"Feynman propagator":
\begin{center}
{$K(b,\ a)\ =\ {\int}_{a}^{b}\ e^{{\frac {i}{{\hbar}}}\ S_{c{\ell}}(b,\
a)}\ D\ x(t)$\ ,\ \ \ \ \ (4.1)}
\end{center}
with:
\begin{center}
{$S_{c{\ell}}(b,\ a)\ =\ {\int}_{t_{a}}^{t_{b}}\ L\ (x,\ {\dot {x}},\
t)\ dt$\ ,\ \ \ \ \ (4.2)}
\end{center}
where $L(x,\ {\dot {x}},\ t)$ is the Lagrangean and $D\ x(t)$ is the
Feynman's Measurement. It indicates that we must perform the integration
taking into account all the ways connecting the states ${\mid}\ a\ >$
and ${\mid}\ b\ >$.
\par
Note that the integral which defines $K(b,\ a)$\ is called "path
integral" or "Feynman integral" and that the Schr\"{o}dinger
wavefunction ${\psi}(x,\ t)$ of any physical system is given by (we
indicate the initial position and initial time by $x_{o}$ and $t_{o}$,
respectively):\ [9]
\begin{center}
{${\psi}(x,\ t)\ =\ {\int}_{-\ {\infty}}^{+\ {\infty}}\ K(x,\ x_{o},\
t,\ t_{o})\ {\psi}(x_{o},\ t_{o})\ dx_{o}$\ ,\ \ \ \ \ (4.3)}
\end{center}
with the quantum causality condition:
\begin{center}
{${\lim\limits_{t,\ t_{o}\ {\to}\ 0}}\ K(x,\ x_{o},\ t,\ t_{o})\ =\
{\delta}(x\ -\ x_{o})$\ .\ \ \ \ \ (4.4)}
\end{center}
\vspace{0.2cm}
\par
4.2.\ {\bf Calculation of the Feynman-de Broglie-Bohm Propagator
for the Linearized Kostin Equation along a Classical Trajetory}
\vspace{0.2cm}
\par
According to Section 3, the wavefunction ${\psi}(x,\ t)$ that was
named wave packet of the of the linearized Kostin equation along
a classical trajetory, can be written as [see (3.22)]:
\begin{center}
{${\psi}(x,\ t)\ =\ [2\ {\pi}\ a^{2}(t)]^{-\ 1/4}\ exp\ {\Bigg
{[}}\ {\frac {i\ m}{2\ {\hbar}}}\ {\Big {(}}\ {\frac {{\dot
{a}}(t)}{a(t)}}\ - {\frac {1}{4\ a^{2}(t)}}\ {\Big {)}}\ [x\ -\
q(t)]^{2}\ {\Bigg {]}}\ {\times}$}
\end{center}
\begin{center}
{${\times}\ exp\ {\Bigg {[}}\ {\frac {i\ m\ {\dot {q}}(t)}{{\hbar}}}\ [x\
-\ q(t)]\ +\ {\frac {i\ m\ v_{o}\ x_{o}}{{\hbar}}}\ {\Big {]}}\ {\times}$}
\end{center}
\begin{center}
{${\times}\ exp\ {\Bigg {[}}\ {\frac {i}{{\hbar}}}\ {\int}_{o}^{t}\ dt'\
{\Bigg {(}}\ {\frac {1}{2}}\ m\ {\dot {q}}^{2}(t')\ +\ {\hbar}\
{\nu}\ [S_{0}(t')\ -\ V[q(t'),\ t']\ -\ {\frac {{\hbar}^{2}}{4\ m\
a^{2}(t')}}\ {\Bigg {)}}\ {\Bigg {]}}$\ .\ \ \ \ \ (4.5)}
\end{center}
where [see (3.16,17)]:
\begin{center}
{${\ddot {q}}(t)\ +\ {\nu}\ {\dot {q}}(t)\ +\ {\frac {1}{m}}\
V'[q(t),\ t]\ =\ 0$\ ,\ \ \ \ \ (4.6)}
\end{center}
\begin{center}
{${\ddot {a}}(t)\ +\ {\nu}\ {\dot {a}}(t)\ +\ {\frac {1}{m}}\
V"[q(t),\ t]\ =\ {\frac {{\hbar}^{2}}{4\ m^{2}\ a(t)^{3}}}$\ .\ \ \ \ \
(4.7)}
\end{center}
where the following initial conditions were obeyed [see (3.18a-d)]:
\begin{center}
{$q(0)\ =\ x_{o}\ ,\ \ \ {\dot {q}}(0)\ =\ v_{o}\ ,\ \ \ a(0)\
=\ a_{o}\ ,\ \ \ {\dot {a}}(0)\ =\ b_{o}$\ .\ \ \ \ \ \
(4.8a-d)}
\end{center}
\par
Therefore, considering (4.3), the Feynman-de Broglie-Bohm propagator
will be calculated using (4.5), in which we will put with no loss of
generality, $t_{o}\ =\ 0$. Thus:
\begin{center}
{${\psi}(x,\ t)\ =\ {\int}_{-\ {\infty}}^{+\ {\infty}}\ K(x,\ x_{o},\
t)\ {\psi}(x_{o},\ 0)\ dx_{o}$\ .\ \ \ \ \ (4.9)}
\end{center}
\par
Let us initially define the normalized quantity:
\begin{center}
{${\Phi}(v_{o},\ x,\ t)\ =\ (2\ {\pi}\ a_{o}^{2})^{1/4}\ {\psi}(v_{o},\
x,\ t)$\ ,\ \ \ \ \ (4.10)}
\end{center}
which satisfies the following completeness relation:\ [10]
\begin{center}
{${\int}_{-\ {\infty}}^{+\ {\infty}}\ dv_{o}\ {\Phi}^{*}(v_{o},\ x,\
t)\ {\Phi}(v_{o},\ x',\ t)\ =\ ({\frac {2\ {\pi}\ {\hbar}}{m}})\
{\delta}(x\ -\ x')$\ .\ \ \ \ \ (4.11)}
\end{center}
\par
Considering (2.4a), (2.9a) and (4.9), we get:
\begin{center}
{${\Phi}^{*}(v_{o},\ x,\ t)\ {\psi}(v_{o},\ x,\ t)\ =$}
\end{center}
\begin{center}
{$=\ (2\ {\pi}\ a_{o}^{2})^{1/4}\ {\psi}^{*}(v_{o},\ x,\ t)\
{\psi}(v_{o},\ x,\ t)\ =\ (2\ {\pi}\ a_{o}^{2})^{1/4}\ {\rho}(v_{o},\
x,\ t)\ \ \ {\to}$}
\end{center}
\begin{center}
{${\rho}(v_{o},\ x,\ t)\ =\ (2\ {\pi}\ a_{o}^{2})^{-\ 1/4}\
{\Phi}^{*}(v_{o},\ x,\ t)\ {\psi}(v_{o},\ x,\ t)$\ .\ \ \ \ \
(4.12)}
\end{center}
\par
On the other side, substituting (4.12) into (2.11), integrating the
result and using (3.1) and (4.10) results [remembering
that ${\psi}^{*}\ {\psi}({\pm}\ {\infty})\ \ \ {\to}\ \ \ 0$]:
\begin{center}
{${\frac {{\partial}({\Phi}^{*}\ {\psi})}{{\partial}t}}\ +\ {\frac
{{\partial}({\Phi}^{*}\ {\psi}\ v_{qu})}{{\partial}x}}\ =\ 0\ \ \ {\to}$}
\end{center}
\begin{center}
{${\frac {{\partial}}{{\partial}t}}\ {\int}_{-\ {\infty}}^{+\ {\infty}}\
dx\ {\Phi}^{*}\ {\psi}\ +\ ({\Phi}^{*}\ {\psi}\ v_{qu}){\mid}_{-\
{\infty}}^{+\ {\infty}}\ =$}
\end{center}
\begin{center}
{$=\ {\frac {{\partial}}{{\partial}t}}\ {\int}_{-\ {\infty}}^{+\ {\infty}}\
dx\ {\Phi}^{*}\ {\psi}\ +\ (2\ {\pi}\ a_{o}^{2})^{1/4}\ ({\psi}^{*}\
{\psi}\ v_{qu}){\mid}_{-\ {\infty}}^{+\ {\infty}}\ =\ 0\ \ \ {\to}$}
\end{center}
\begin{center}
{${\frac {{\partial}}{{\partial}t}}\ {\int}_{-\ {\infty}}^{+\
{\infty}}\ dx\ {\Phi}^{*}\ {\psi}\ =\ 0$\ .\ \ \ \ \ (4.13)}
\end{center}
\par
Eq. (4.13) shows that the integration is time independent.
Consequently:
\begin{center}
{${\int}_{-\ {\infty}}^{+\ {\infty}}\ dx'\ {\Phi}^{*}(v_{o},\ x',\ t)\
{\psi}(x',\ t)\ =\ {\int}_{-\ {\infty}}^{+\ {\infty}}\ dx_{o}\
{\Phi}^{*}(v_{o},\ x_{o},\ 0)\ {\psi}(x_{o},\ 0)$\ .\ \ \ \ \ (4.14)}
\end{center}
\par
Multiplying (4.14) by ${\Phi}(v_{o},\ x,\ t)$ and integrating over
$v_{o}$ and using (4.11), we obtain [remembering
that ${\int}_{-\ {\infty}}^{+\ {\infty}}\ dx'\ f(x')\ {\delta}(x' -\
x)\ = f(x)$]:
\begin{center}
{${\int}_{-\ {\infty}}^{+\ {\infty}}\ {\int}_{-\ {\infty}}^{+\
{\infty}}\ dv_{o}\ dx'\ {\Phi}(v_{o},\ x,\ t)\ {\Phi}^{*}(v_{o},\ x',\ t)\
{\psi}(x',\ t)$\ =}
\end{center}
\begin{center}
{=\ ${\int}_{-\ {\infty}}^{+\ {\infty}}\ {\int}_{-\
{\infty}}^{+\ {\infty}}\ dv_{o}\ dx_{o}\ {\Phi}(v_{o},\ x,\ t)\
{\Phi}^{*}(v_{o},\ x_{o},\ 0)\ {\psi}(x_{o},\ 0)\ \ \ {\to}$}
\end{center}
\begin{center}
{${\int}_{-\ {\infty}}^{+\ {\infty}}\ dx'\ ({\frac {2\ {\pi}\
{\hbar}}{m}})\ {\delta}(x'\ -\ x)\ {\psi}(x',\ t)\ =\ ({\frac {2\ {\pi}\
{\hbar}}{m}})\ {\psi}(x,\ t)$\ =}
\end{center}
\begin{center}
{=\ ${\int}_{-\ {\infty}}^{+\ {\infty}}\ {\int}_{-\ {\infty}}^{+\
{\infty}}\ dv_{o}\ dx_{o}\ {\Phi}(v_{o},\ x,\ t)\ {\Phi}^{*}(v_{o},\
x_{o},\ 0)\ {\psi}(x_{o},\ 0)\ \ \ {\to}$}
\end{center}
\begin{center}
{${\psi}(x,\ t)\ =\ {\int}_{-\ {\infty}}^{+\ {\infty}}\ {\Big {[}}\
({\frac {m}{2\ {\pi}\ {\hbar}}})\ {\int}_{-\ {\infty}}^{+\ {\infty}}\
dv_{o}\ {\Phi}(v_{o},\ x,\ t)\ {\times}$}
\end{center}
\begin{center}
{${\times}\ {\Phi}^{*}(v_{o},\ x_{o},\ 0)\ {\Big {]}}\ {\psi}(x_{o},\
0)\ dx_{o}$\ .\ \ \ \ \ (4.15)}
\end{center}
\par
Comparing (4.9) and (4.15), we have:
\begin{center}
{$K(x,\ x_{o},\ t)\ =\ {\frac {m}{2\ {\pi}\ {\hbar}}}\ {\int}_{-\
{\infty}}^{+\ {\infty}}\ dv_{o}\ {\Phi}(v_{o},\ x,\ t)\
{\Phi}^{*}(v_{o},\ x_{o},\ 0)$\ .\ \ \ \ \ (4.16)}
\end{center}
\par
Substituting (4.5) and (4.10) into (4.16), we finally obtain the
Feynman-de Broglie-Bohm Propagator of the linearized Kostin
equation along a classical trajetory, remembering that
${\Phi}^{*}(v_{o},\ x_{o},\ 0)\ =\ exp\ (-\ {\frac {i\ m\ v_{o}\
x_{o}}{{\hbar}}})$]:
\begin{center}
{$K(x,\ x_{o};\ t)\ =\ {\frac {m}{2\ {\pi}\ {\hbar}}}\ {\int}_{-\
{\infty}}^{+\ {\infty}}\ dv_{o}\ {\sqrt {{\frac
{a_{o}}{a(t)}}}}\ {\times}$}
\end{center}
\begin{center}
{${\times}\ exp\ {\Big {[}}\ {\Big {(}}\ {\frac {i\ m\ {\dot
{a}}(t)}{2\ {\hbar}\ a(t)}}\ -\ {\frac {1}{4\
a^{2}(t)}}\ {\Big {)}}\ [x\ -\ q(t)]^{2}\ +\ {\frac {i\ m\ {\dot
{q}}(t)}{{\hbar}}}\ [x\ -\ q(t)]\ {\Big {]}}\ {\times}$}
\end{center}
\begin{center}
{${\times}\ exp\ {\Big {[}}\ {\frac {i}{{\hbar}}}\
{\int}_{o}^{t}\ dt'\ {\Big {(}}\ {\frac {1}{2}}\ m\ {\dot
{q}}^{2}(t')\ +\ 2\ {\hbar}\ {\nu}\ [S_{0}(t')]\ -\ {\frac
{{\hbar}^{2}}{4\ m\ a^{2}(t')}}\ {\Big {)}}\ {\Big {]}}$\ ,\ \ \
(4.17)}
\end{center}
where $q(t)$ and $a(t)$ are solutions of the (4.6,\ 7) differential
equations.
\par
Finally, it is important to note that putting ${\alpha}\ =\ 0$ into
(3.14), (3.15) and (4.17) we obtain the free particle Feynman
propagator.\ [7,\ 9]
\par
\begin{center}
{{\bf NOTES AND REFERENCES}}
\end{center}
\par
1.\ KOSTIN, M. D. {\it Journal of Chemical Physics 57}, p. 3539 (1972).
\par
2.\ MADELUNG, E. {\it Zeitschrift f\"{u}r Physik 40}, 322 (1926).
\par
3.\ BOHM, D. {\it Physical Review 85}, 166 (1952).
\par
4.\ BASSALO, J. M. F., ALENCAR, P. T. S., CATTANI, M. S. D. e
NASSAR, A. B. {\it T\'opicos da Mec\^anica Qu\^antica de de
Broglie-Bohm}, EDUFPA (2003).
\par
5.\ BASSALO, J. M. F., ALENCAR, P. T. S., SILVA, D. G., NASSAR, A. B.
and CATTANI, M. {\it arXiv:0905.4280v1}\ [quant-ph]\ 26\ May\
2009.
\par
6.\ BASSALO, J. M. F., ALENCAR, P. T. S., SILVA, D. G., NASSAR, A. B.
and CATTANI, M. {\it arXiv:1004.1416v1}\ [quant-ph]\ 10\ April\
2010.
\par
7. NASSAR, A. B., BASSALO, J. M. F., ALENCAR, P. T. S., CANCELA, L. S.
G. and CATTANI, M. {\it Physical Review 56E}, 1230 (1997).
\par
8.\ FEYNMAN, R. P. {\it Reviews of Modern Physics 20}, 367 (1948).
\par
9.\ FEYNMAN, R. P. and HIBBS, A. R. {\it Quantum Mechanics and Path
Integrals}, McGraw-Hill Book Company (1965).
\par
10.\ BERNSTEIN, I. B. {\it Physical Review A32}, 1 (1985).
\end{document}